\documentstyle[preprint,prl,aps]{revtex}
\begin{document}

\title{The Absence of the Fractional Quantum Hall Effect at High Landau Levels}
\author{M. M. Fogler and A. A. Koulakov}

\address{Theoretical Physics Institute, University of Minnesota,
116 Church St. Southeast, Minneapolis, Minnesota 55455}

\maketitle
\draft

%\psdraft
%\psfig{figure=stamp.ps,rheight=0pt,rwidth=0pt,bbllx=170pt,bblly=580pt,bburx=475pt,bbury=608pt,angle=30}

\begin{abstract}

We compare the energies of the Laughlin liquid and a charge density
wave in a weak magnetic field
for the upper Landau level filling factors $\nu_N = 1/3$ and $1/5$.
The charge density wave period has been optimized
and was found to be $\simeq 3R_c$, where $R_c$ is the cyclotron radius.
We conclude that the optimal charge density wave is more energetically
preferable than the Laughlin liquid for the Landau level numbers
$N \ge 2$ at $\nu_N = 1/3$ and for $N \ge 3$ at $\nu_N = 1/5$.
This implies that the $1/3$ fractional quantum Hall effect
cannot be observed for $N \ge 2$, in agreement with the experiment.

\end{abstract}
\pacs{PACS numbers: 73.40.Hm}

The fractional quantum Hall effect (FQHE) was first discovered at the lowest
Landau level (LL)\cite{Prange}. This remarkable phenomenon
occuring at certain unique values of the filling factor
$\nu = 1/3, 1/5, \ldots$ has been associated with the formation
of a uniform incompressible quantum state, or the Laughlin
liquid\cite{Laughlin}.
The traditional alternative to the Laughlin liquid is a charge density wave (CDW),
which does not exhibit the FQHE.
The FQHE occurs because
the Laughlin liquid is lower in energy than the optimal CDW,
which at the lowest LL has the same spacial periodicity
as the triangular Wigner crystal~\cite{Yoshioka}.

Later, the FQHE was found at the first excited LL also.
The theoretical work, driven by this discovery~\cite{Girvin MacDonald},
addressed the question of what exactly ground state,
liquid or crystalline, is formed at high LL's?
The liquid state at the $N$-th LL was defined as follows:
\begin{equation}
\left| \Psi _L^N\right\rangle =\prod
\limits_i\frac{\left( a_i^{\dagger }\right) ^N}
{\sqrt{N!}}\left| \Psi _L^0\right\rangle
\end{equation}
Here $a_i^{\dagger }$ is the {\em inter}-LL ladder operator,
raising the $i$-th electron to the next LL, and
$\left| \Psi _L^0\right\rangle$ is the Laughlin state at the lowest LL.
As a crystalline state Ref.~\onlinecite{Girvin MacDonald}
continues using the Wigner crystal as well as at the lowest LL.
However, a recent more elaborate investigation
of the CDW state\cite{Short} shows that the optimum one should
have a period of the order of $R_c$, the cyclotron radius.
The formation of the CDW with this period, which is a characteristic
spread of electron wave functions,
enables the system to reach a lower value of the interaction energy.
Below we compare the energies of this optimal CDW
and the Laughlin liquid. We show that the crystalline ground state
rules out the FQH states at high LL's.

Our calculation is based on the following model\cite{Aleiner Glazman}.
We explicitly consider only the electrons at the upper LL,
which is assumed to be spin-polarized.
All the other LL's are completely filled.
The role of these lower LL's is reduced to the screening
of the Coulomb interaction among the electrons at the upper LL.
The screening is accounted for by means of the dielectric
function~\cite{Aleiner Glazman}:
\begin{equation}
\begin{array}{cc}
{\displaystyle
\epsilon \left( q\right) =1 + v \left( q\right) \Pi \left( q\right),} \\ \\
{\displaystyle
\Pi \left( q\right) =\frac {2}{\pi l^2} \sum
\limits_{\begin{array}c {\scriptstyle m < N}  \\ {\scriptstyle N \le n} \end{array}}
\frac{ \left( -1\right) ^{n-m} }{\hbar
\omega _c^{}\left( n-m\right) }F_{nm}\left( q\right) F_{mn}\left( q\right),} \\
{\displaystyle F_{nm}=L_m^{n-m}\left( \frac{q^2l^2}2\right)
e^{-\frac{q^2l^2}4},}
\end{array}
\label{Epsilon}
\end{equation}
where $l$ is the magnetic length, $\kappa$ is the bare dielectric constant,
$v\left( q\right) =2\pi e^2/\kappa q$ is the Coulomb potential,
$L_n^m(x)$ is the Laguerre polynomial, and $\omega _c^{}$
is the cyclotron frequency.
This dielectic function tends to unity in the limits $q \to 0$ and
$q \to \infty$ and reaches its largest value of $1 + \sqrt{2} \, Nr_s$
at $q \sim R_c^{-1}$.
Here $r_s^{}=\sqrt{2}/k_{\rm F}^{}a_{\rm B}^{}$,
with $k_{\rm F}^{}$ being the Fermi wave vector, and $a_{\rm B}^{}$ being
the effective Bohr radius.
This model correctly renders the
low energy physics of the system in the limit $r_s \ll 1$ and $Nr_s \gg 1$
[Ref.~\onlinecite{Aleiner Glazman}].
Moreover, the results obtained
within the framework of this model remain correct to the leading order in $r_s$
even for $Nr_s \ll 1$. In the latter limit $\epsilon \left( q\right) \simeq 1$,
which is consistent with the fact that the LL mixing can be ignored completely.

Let us now describe the CDW state at the upper LL.
In the limit of a weak magnetic field ($N \gg 1$)
a simple quasiclassical picture can be given.
In this case electrons can be viewed
as classical particles rotating in cyclotron orbits.
The only constraint imposed by the Landau quantization is that
the concentration of the centers of the cyclotron
circles at any point does not exceed $1/2\pi l^2$.
One can fill in a disk with these centers at their maximum concentration
(see Fig.~\ref{fig1}).
We call this disk a bubble
and the triangular crystal built out of such disks a bubble phase.
It is shown in Ref.~\onlinecite{Short} that the optimum number
of electrons in a bubble  is
\begin{equation}
\tilde{M} \simeq 3\nu_N^{}N,
\label{ElNum}
\end{equation}
which corresponds to the separation $\simeq 3R_c$ between nearest bubbles.
The nonuniform distribution of guiding centers with such a period,
chosen in accordance with the form factor of the electron wave function,
does not create significant variations of the charge density.
Hence, the electrostatic (Hartree)
energy of the system does not increase too much. However, the
exchange interaction in this ferromagnetic state favors
an increase of overlap among the wave functions
and thus the most compact arrangement of the guiding centers.

For the comparison of different trial states
we use the cohesive energy, which is the energy per electron
relative to the uniform state formed at high temperature.
The cohesive energy of the CDW state has been shown to be
$E_{\rm coh}^{\rm CDW} = - r_s^{}\hbar \omega _c^{}\propto - B$
[Ref.~\onlinecite{Short}],
which is of the same order of magnitude as the exchange-enhanced
spin splitting~\cite{Aleiner Glazman}.

As for the cohesive energy of the Laughlin liquid,
arguments can be given~\cite{Aleiner Glazman} that it is very
close to the cohesive energy of the Wigner crystal. The lower bound
for the latter was estimated in Ref.~\onlinecite{Short} to be
$E_{\rm coh}^{\rm L}\sim - \hbar \omega _c^{}/\sqrt{N} \propto - B^{3/2}$.
In the limit of small magnetic field the CDW should obviously
be more energetically preferable, ruling out the FQH states.

It is interesting to know at what LL the transition from
liquid to the crystalline ground state occurs?
To answer this question one cannot use the quasiclassical
approach and the CDW should be defined more accurately.
We do so in several steps.
First, we introduce the wave function of one bubble
consisting of $M$ electrons at the {\em lowest} LL:
\begin{equation}
\displaystyle
\Psi _0^{}\left\{ \bbox{r}_k^{}\right\}
=\left\{ \prod\limits_{1\leq i<j\leq M^{}}
\left( z_i-z_j\right) \right\} \exp \left( -\sum\limits_{i=1}^{M^{}}
\frac{\left| z_i\right| ^2}{4l^2}\right).
\label{Bubble0}
\end{equation}
Here $z_j = x_j + iy_j$ is the complex coordinate of the $j$-th electron.
Second, we define the wave function of a bubble at the $N$-th LL
centered at point $R$ by raising every electron onto this LL
and shifting its position by the
magnetic translation operator\cite{Kivelson}
\begin{equation}
\displaystyle
\Psi _{\bbox{R}}^{}\left\{ \bbox{r}_k^{}\right\} =\prod\limits_{i=1}^{M}
\frac{\left( a_i^{\dagger } \right) ^N}{\sqrt{N!}}
\exp \left( \frac{b_i^{\dagger }\bar{R}-b_i^{}R}{l\sqrt{2}}\right)
\Psi _0^{}\left\{ \bbox{r}_k^{}\right\},
\end{equation}
where $b_i$ is an {\em intra}-LL ladder operator.
To finally obtain the wave function of the CDW
we build an antisymmetric combination of the bubbles centered
at the triangular lattice sites $\bbox{R}_l^{}$
\begin{equation}
\displaystyle
\Psi _{\rm CDW}=\sum\limits_P{\rm sign}\left( P\right)
\prod\limits_{\bbox{R}_l}\Psi _{\bbox{R}_l^{}}^{}
\left\{ P\left( \bbox{r}_k^{}\right) \right\}.
\label{CDWWF}
\end{equation}
Here $P$ are the permutations of electrons between bubbles.
For the case $M=1$ this trial state coinsides with the
Maki-Zotos ansatz wave function for the Wigner crystal~\cite{Maki Zotos}.
It can be easily seen that $\Psi _{\rm CDW}$ is of the Fock type,
and that the overlap between the wave functions of different bubbles
is negligible.

To show that $\Psi _{\rm CDW}$ matches our earlier quasiclassical picture
we introduce the guiding center density operator:
\begin{equation}
\hat{\nu}\left( \bbox{r}\right) = 2\pi l^2 \sum\limits_i^{}
\delta \left( \bbox{r}-\hat{\bbox{R}}_i^{}\right).
\end{equation}
The summation here is carried over the electrons at the considered LL and
$\hat{\bbox{R}}_i^{}=\bbox{r}_i^{}+\frac{l^2}\hbar \bbox{z}\times
\hat{\bbox{P}_i}$
is the guiding center operator, with $\hat{\bbox{P}_i}$ being
the canonical momentum of the $i$-th particle.
The average of this operator has a physical meaning
of the guiding center density~\cite{GCDensity}. It can be shown that for
the state defined by Eq.~(\ref{CDWWF})
\begin{equation}
\begin{array}{cc}
{\displaystyle
\left\langle \hat{\nu} \left( q\right) \right\rangle
=\frac {\nu _N^{}A}M
F_{M,M-1}\left( q\right)
\approx 2\nu _N^{}A
\frac{J_1\left( ql\sqrt{2M}\right) }{ql\sqrt{2M}},} \\ \\
{\displaystyle
q\ll \frac{\sqrt{M}}l,}
\end{array}
\label{OrderParameter}
\end{equation}
which is just the Fourier transform of a uniform disk
with the radius $l\sqrt{2M}$. The last equation
in (\ref{OrderParameter}) follows from the asymptotic formula
for the Laguerre polynomials~\cite{Erdelyi}.

The cohesive energy of the CDW can be calculated in the same way
as it has been done for the Wigner crystal\cite{Yoshioka}:
\begin{equation}
E_{\rm coh}^{\rm CDW}=\frac 1{2\nu _N^{}}\sum\limits_{q\neq 0}^{}
u^{}_{\rm HF}\left( q\right)
\left| \frac { \left\langle \hat{\nu} \left( q\right) \right\rangle }
{A} \right| ^2
\label{Ecoh}
\end{equation}
The summation in Eq.~(\ref{Ecoh}) is carried over the reciprocal vectors
of the triangular lattice.
The Hartree-Fock interaction potential 
$u^{}_{\rm HF}\left( q\right)$ is defined
in the same way as in Ref.~\onlinecite{Short}:
\begin{equation}
\begin{array}{cc}
{\displaystyle u_{\rm HF}^{}\left( q\right) =u_{\rm H}^{}\left( q\right)
-u_{\rm ex}^{}\left( q\right),} \\ \\
{\displaystyle u_{\rm H}^{}\left( q\right) =\frac{v\left( q\right) }
{2\pi l^2 \epsilon \left( q\right) }F_{NN}\left( q\right),} \\ \\
{\displaystyle u_{\rm ex}^{}\left( q\right) =2\pi l^2\int \frac{d^2q^{\prime }}
{\left( 2\pi \right) ^2}e^{i\bbox{qq^{\prime }}l^2}u_{\rm H}^{}
\left( q^{\prime }\right).}
\end{array}
\end{equation}
Using Eqs.~(\ref{OrderParameter},\ref{Ecoh}), the cohesive energy
for any given $\nu_N$ can be
calculated numericaly.
The result is, of course, different for different $M$ (see Fig.~\ref{fig2}).
Therefore, one has to find $\tilde{M}$ corresponding to the lowest energy.
The energies of the CDW optimized in this way
are summarized in Tables~\ref{Table1} and ~\ref{Table2}.
In Table~\ref{Table1} we present the
results for the case $Nr_s \ll 1$, when the LL mixing can be
ignored completely, i.e., $\epsilon (q) \equiv 1$. Table~\ref{Table2}
shows the results for the practically important case $r_s = \sqrt{2}$.
One can see that the optimum number of electrons per bubble is the
same both with and without the screening and is
in perfect agreement with Eq.~(\ref{ElNum}).

The above results have been tested by the self-consistent Hartree-Fock 
procedure,
similar to that described in Ref.~\onlinecite{Yoshioka}.
Starting from the initial approximation given by wave function
~(\ref{CDWWF}), this procedure finds the optimal set of
$\left\langle \hat{\nu} \left( q\right) \right\rangle$
for a given periodicity of the CDW.
The obtained corrections are of the order of $10^{-5}r_s^{} \hbar \omega _c$
and thus do not affect the significant
digits displayed in Tables~\ref{Table1}~and~\ref{Table2}.
We associate the corrections
with a slight nonorthogonality of the wave functions of different
bubbles.

Let us now discuss the Laughlin liquid at high LL's.
The interaction energy per electron can be calculated
using the density-density correlation function
\begin{equation}
h_N^{}\left( \bbox{r}\right) \equiv
\frac{\left\langle \hat{\rho}_N\left( \bbox{r}\right)
\hat{\rho}_N\left( 0\right) \right\rangle -\left\langle
\hat{\rho}_N\right\rangle ^2}{\left\langle \hat{\rho}_N\right\rangle }
\end{equation}
where $\hat\rho_N(\bbox{r})$ is the
projection of the density operator onto the $N$-th LL.
This can be most effectively done in the Fourier space because
$h_N \left( q \right)$ is very simply related to $h_0 \left( q \right)$
(the correlation function for $N = 0$), obtained earlier by Monte-Carlo
simulations~\cite{Girvin MacDonald}:
\begin{equation}
\begin{array}{c}
{\displaystyle
h_N\left( q\right)
=h_0\left( q\right) \left[ L_N\left( \frac{q^2l^2}2\right) \right] ^2,} \\
{\displaystyle
E_{\rm cor}^{\rm L}
=\frac 12\int\limits_{}^{}\frac{d^2q}{\left( 2\pi \right) ^2}
\frac{v\left( q\right) }
{\epsilon \left( q\right) }h_N\left( q\right)
.}
\end{array}
\end{equation}
The calculations are greatly simplified by virtue of the approximation
formula for $h_0$ given in
Ref.~\onlinecite{Mac Girv Platz}.

The cohesive energy per electron is then, according to the definition
\begin{equation}
E_{\rm coh}=E_{\rm cor}-E^{\rm UEL}
=E_{\rm cor}+\frac {\nu_N}2 u_{\rm ex}\left( q=0\right),
\end{equation}
where $E^{\rm UEL}$ is the interaction energy per particle
in the uniform uncorrelated
electron liquid formed at high temperature. The results of the numerical
evaluation of these energies are also listed in Tables~\ref{Table1}~
and~\ref{Table2}.

Compare now the energies of the Laughlin liquid and the CDW.
As one can see,
at $N=0$ and $N=1$ the Laughlin liquid is more energetically preferable.
At large $N$ however the CDW wins. The transition to the bubble state
both with and without screening occurs at $N=2$ for $\nu _N =1/3$
and at $N=3$ for $\nu _N =1/5$.
The difference in the energies of these two states at $N=2$ and $\nu_N=1/5$
is very small. It should be noted in this connection that
the trial state~(\ref{CDWWF}), being compressible in nature,
can be further improved by introducing the magnetophonon
correlations~\cite{Lam_Girvin}. Another possibility
arises when the system is subjected to an external impurity potential.
The CDW can lower its energy by accomodating
to this potential, while the incompressible $\nu_N = 1/5$
liquid state cannot~\cite{ExtDisorder}.
Hence, the question of what exactly phase dominates at $\nu = 4\frac 15$
remains open.

%
%We emphasize that
%the results of Table ~\ref{Table1}, corresponding to the limit $Nr_s \ll 1$
%are exact to the leading order in $r_s$.
%

One can suggest the following interpretation of
our results considering the zero-point vibrations of the lattice.
Electrons at any $N$ form the Wigner crystal
when the upper LL filling is so small that the average
distance between electrons
is larger than the characteristic spread of electron wave functions $R_c$.
The amplitude of the zero-point vibrations
of the crystal is mostly determined by the magnetic barrier
since the interaction energy per electron
is smaller than the cyclotron gap for the cases of interest.
Hence this amplitude is $\lesssim l$.
For the crystalline state to be stable this amplitude should be
small compared to the lattice constant (the Lindemann criterion).
Now an important difference between the low and high LL's arises.
At low LL's an increase of the electron density
reduces the lattice constant. At some value of $\nu_N$ it becomes of the
order of $l$ and the crystal melts into the Laughlin liquid.
At high LL's, however, the lattice constant does not change
as one increases the LL filling but remains of the order of $R_c \gg l$.
Hence CDW cannot be melted by quantum fluctuations at high LL's.

In conclusion, we have compared the energies of the Laughlin liquid
and the CDW with
the optimized period ($\sim R_c$) at the upper LL filling factors
$\nu_N = 1/3$ and $1/5$. We found that
the $1/3$ liquid state is unstable for $N \ge 2$, while
the $1/5$ state loses to the CDW at $N \ge 3$.
Our result implies that the $1/3$ fractional quantum Hall effect cannot
be observed at filling factors $\nu > 4$.
This conclusion is in agreement
with the existing experimental data.
The difference between the energies of the CDW and the Laughlin liquid
at $N=2$ and $\nu_N = 1/5$ is so small that more work is needed
to distinguish them unambiguously.
The authors are grateful to B.~I.~Shklovskii for the inspiration  for
this work
and for numerous helpful suggestions.
This work is supported by NSF under Grant DMR-9321417.

%
% Fig 1
%
\begin{figure}
\centerline{\caption{
The quasiclassical image of the bubble phase.
(a) Top view.
The bubbles (dark circles) are the places where
the accumulation of the guiding centers occurs.
(b) The enlarged view of one bubble.
The dark region shows the guiding center density $\nu (x,y)/2\pi l^2$,
while the toroidal figure illustrates the charge density distribution
$\rho_N (x,y)$ around the bubble. Half of the charge density is removed.
This charge density is created by electrons moving
in the cyclotron orbits centered inside the bubble.
One of such cyclotron orbits is shown by the arc with the arrow.
\label{fig1}
}}
\end{figure}

%
% Fig 2
%
\begin{figure}
\centerline{\caption{
The cohesive energy of the CDW as a function of $\nu_ N$ for different
numbers of electrons in a bubble $M$.
The calculations are made for $N=5$ and $r_s=\protect \sqrt{2}$.
The crosses show the Laughlin liquid energies.
\label{fig2}
}}
\end{figure}

\begin{table}

\centerline{
\begin{minipage}[t]{4in}
\renewcommand{\arraystretch}{1.2}
\begin{tabular}{@{}c@{\hspace{5mm}}c@{\hspace{5mm}}c@{\hspace{5mm}}c@{\hspace{5mm}}c@{\hspace{5mm}} c@{}}
\multicolumn{6}{c}{\underline{$\nu _N=1/3$}} \\
$N$	& $\tilde{M}^{}$	& ${\rm E^{UEL}}$	& ${\rm E_{coh}^{L}}$	& ${\rm E_{coh}^{CDW}}$ &$\delta {\rm E/E_{coh}^{CDW}}$ 	\\ \hline
0	& 1		&-0.0853		&-0.0819		&-0.0733		&-11.7\%	\\
1	& 1		&-0.1692		&-0.1824		&-0.1726		&-5.6\%		\\
2	& 2		&-0.1970		&-0.1939		&-0.2163		&10.3\%		\\
3	& 3		&-0.2135		&-0.2056		&-0.2433		&15.5\%		\\
4	& 4		&-0.2251		&-0.2228		&-0.2480		&15.1\%		\\
5	& 5		&-0.2341		&-0.2367		&-0.2767		&14.5\%		\\
\hline \hline
\multicolumn{6}{c}{\underline{$\nu _N=1/5$}} \\
$N$	& $\tilde{M}^{}$	& ${\rm E^{UEL}}$	& ${\rm E_{coh}^{L}}$	& ${\rm E_{coh}^{CDW}}$ &$\delta {\rm E/E_{coh}^{CDW}}$
\\ \hline
0	& 1		&-0.0396 		&-0.0639		&-0.0622			&-2.7\%	\\
1	& 1		&-0.0986 		&-0.2109		&-0.2043			&-3.2\%		\\
2	& 1		&-0.1164		&-0.2465		&-0.2454			&-0.5\%		\\
3	& 2		&-0.1267 		&-0.2474		&-0.2811			&12.0\%		\\
4	& 2		&-0.1340		&-0.2538		&-0.2990			&15.1\%		\\
5	& 3		&-0.1395		&-0.2562		&-0.3187			&19.6\%		\\
\end{tabular}
\end{minipage}
}

\vspace{0.2in}
\centerline{
\caption{The cohesive energies of the Laughlin liquid ${\rm E_{coh}^{L}}$
and the CDW ${\rm E_{coh}^{CDW}}$
in the limit $Nr_s \to 0$ for different $N$.
$\tilde{M}$ is the optimum number of electrons per bubble.
All the energies are given in the units of $r_s \hbar \omega _c$.
The right column shows their relative difference:
$\left( {\rm E_{coh}^{L}} - {\rm E_{coh}^{CDW}} \right) / {\rm E_{coh}^{CDW}}$.
The energy per electron in the uniform uncorrelated state
${\rm E^{UEL}}$ is provided for reference.
\label{Table1}
}
}
\end{table}

\begin{table}
\renewcommand{\arraystretch}{1.2}

\centerline{
\begin{minipage}[t]{4in}
\begin{tabular}{@{}c@{\hspace{5mm}}c@{\hspace{5mm}}c@{\hspace{5mm}}c@{\hspace{5mm}}c@{\hspace{5mm}} c@{}}
\multicolumn{6}{c}{\underline{$\nu _N=1/3$}} \\
$N$	& $\tilde{M}^{}$	& ${\rm E^{UEL}}$	& ${\rm E_{coh}^{L}}$	& ${\rm E_{coh}^{CDW}}$ &$\delta {\rm E/E_{coh}^{CDW}}$ 	\\ \hline
0	& 1		&-0.1206		&-0.1158		&-0.1037		&-11.7\%	\\
1	& 1		&-0.1297		&-0.1520		&-0.1424		&-6.7\%		\\
2	& 2		&-0.1136		&-0.1141		&-0.1188		&4.0\%		\\
3	& 3		&-0.1034		&-0.0948		&-0.1018		&6.9\%		\\
4	& 4		&-0.0965		&-0.0822		&-0.0896		&8.2\%		\\
5	& 5		&-0.0914		&-0.0743		&-0.0805		&7.7\%		\\
\hline \hline
\multicolumn{6}{c}{\underline{$\nu _N=1/5$}} \\
$N$	& $\tilde{M}^{}$	& ${\rm E^{UEL}}$ & ${\rm E_{coh}^{L}}$	& ${\rm E_{coh}^{CDW}}$ &$\delta {\rm E/E_{coh}^{CDW}}$
\\ \hline
0	& 1		&-0.0560 	&-0.0903	&-0.0880	&-2.7\%	\\
1	& 1		&-0.0765	&-0.1731	&-0.1692	&-2.3\%		\\
2	& 1		&-0.0677	&-0.1423	&-0.1396	&-1.9\%		\\
3	& 2		&-0.0618 	&-0.1143	&-0.1202	&4.9\%		\\
4	& 2		&-0.0577	&-0.0979	&-0.1050	&6.8\%		\\
5	& 3		&-0.0547	&-0.0856	&-0.0946	&9.7\%		\\
\end{tabular}
\end{minipage}
}

\vspace{0.2in}
\centerline{
\caption{Same as Table~\protect \ref{Table1} but
$r_s = \protect{\sqrt{2}}$, which corresponds to the electron density of
$1.6\cdot 10^{11}\rm{cm}^{-2}$ in GaAs-GaAlAs heterostructures.
The energies are now given in the units of $\hbar \omega _c$.
\label{Table2}
}
}
\end{table}

\end{document}